# Blockchain based Proxy Re-Encryption Scheme for Secure IoT Data Sharing

Ahsan Manzoor*, Madhsanka Liyanage†‖, An Braeken‡, Salil S. Kanhere§, Mika Ylianttila¶
*‡§Centre for Wireless Communications, University of Oulu, Finland
‖School of Computer Science, University College Dublin, Ireland
‡Industrial Engineering INDI, Vrije Universiteit Brussel, Belgium
§School of Computer Science and Engineering, University of New South Wales, Australia
{ahsan.manzoor*, madhusanka.liyanage†, mika.ylianttila¶}@oulu.fi, madhusanka@ucd.ie‖, an.braeken@vub.be‡,
salil.kanhere@unsw.edu.au§

***Abstract*—Data is central to the Internet of Things (IoT) ecosystem. Most of the current IoT systems are using centralized cloud-based data sharing systems. Involvement of such third-party service provider requires also trust from both sensor owner and sensor data user. Moreover, the fees need to be paid for their services. To tackle both the scalability and trust issues and to automatize the payments, this paper presents a blockchain based proxy re-encryption scheme. The system stores the IoT data in a distributed cloud after encryption. To share the collected IoT data, the system establishes runtime dynamic smart contracts between the sensor and the data user without the involvement of a trusted third party. It also uses an efficient proxy re-encryption scheme which allows that the data is only visible by the owner and the person present in the smart contract. The proposed system is implemented in an Ethereum based testbed to analyze the performance and security properties.**

## I. Introduction

The Internet of Things (IoT) is an emerging technology which has great technical, social, and economic significance. Current predictions for the impact of IoT are very impressive. With the development of 5G, it is anticipating that 100 billion connected IoT devices will be used by 2025 [1], [2]. It will also have a global economic impact of more than $11 trillion [3] [4].

Data is central to the IoT paradigm. IoT data is collected to serve many different types of applications such as smart home, smart city, wearable, healthcare, smart grid, autonomous vehicles, smart farms, industries and manufacturing, and retail sector [4]–[6]. Therefore, numerous heterogeneous sensors exist to measure a variety of parameters. The collected data from these IoT sensors can be useful for different stakeholders. For instance, air quality measurements are of interest to governmental organizations, application developers and inhabitants of the relevant spaces. However, many challenges arise when organizing this data sharing as these IoT devices, which are typically resource-constrained, require efficient mechanisms to guarantee the data integrity and to enable proper processing and security [7]. Due to the large number of IoT devices, scalable deployment, and maintenance costs [5] should also be taken into account. Currently, almost all the sensor systems upload the data to a centralized cloud and share the sensor data with different stakeholders, who prove access to the cloud storage [8]. The sensors get services from the third-party cloud service provider, such as access control in addition to the data storage. In that case, both sensor and sensor data user have to trust the third-party service provider and also need to pay some fee for their services. In addition, it is needed to establish an agreement between the third-party service provider and sensor data user. Most of these agreements are static and take lots of time and administration to be established [9]. It will result in a significant increase of time before the actual data sharing can be realized [10]. Thus, the current centralized architecture model in IoT systems will struggle to scale up to meet the demands of future IoT systems.

**Our Contribution:** To solve these issues, we propose a novel blockchain based scheme in combination with a proxy re-encryption mechanism to ensure the confidentiality of the data. Here, the advantage of using blockchain mechanisms to sell the sensor measurements with different users is that the corresponding financial transactions are automatically managed through the agreed smart contract, stored at the blockchain. Moreover, the availability and other quality of service requirements from the legal contract between both parties can be automatically applied. Consequently, compared to the business scenario where the data is stored in a cloud-based infrastructure, there is no need for manual verification of the payments and the predefined requirements. Also, disputes on these aspects are completely avoided.

The remainder of this paper has the following structure. Section II gives an overview of related work. The proposed architecture and proxy re-encryption scheme is explained in Section III and IV. Section V discusses the implementation of the proposed scheme. The performance analysis results are presented in Section VI. Finally, Section VII presents our conclusions.

## II. Related work

There exist different studies on the security and privacy of the IoT [11]–[16] and the vast majority of this research work is on understanding and identifying these threats [17]–[21]. Moreover use of blockchain to secure various IoT Platforms were discussed in [22]–[26]. The IoT devices sense, gather and share a large amount of data, thus opening up significant security and privacy concerns. Khan and Salah [27] in their paper have reviewed different security challenges to IoT and identified insecure transferring of IoT data as a high-level security risk. Authors in [28] demonstrated the lack of basic security by hacking off-the-shelf smart home IoT devices.

In 1998, Blaze, Bleumer, and Strauss [29] initially introduced the concept of proxy re-encryption and constructed the first bidirectional proxy re-encryption application. Authors in [30], [31] also propose a similar scheme but it is not dynamic, hence making it unsuitable for cloud data sharing. In [32], a very efficient solution for data storage in the cloud is proposed using a pairing free proxy re-encryption scheme. However, the scheme is not implemented in practice. Although the underlying structure of our proposed scheme is based on it, some important modification like the inclusion of metadata is included to ensure a practical usage of the scheme.

Most of the prior work partly addresses the problem of securely sharing the IoT data. It is nearly impossible to come up with device-embedded security to solve all the security threats to the IoT devices. Limited computing and power resources of IoT also make the execution of complex security algorithms harder on the device. We propose using the combination of a blockchain and a paring free proxy re-encryption scheme to provide a trading platform and to ensure secure transfer of the sensor data to the user.

## III. Proposed Architecture

In this section, we present our new architecture based on the mechanisms of blockchain and re-encryption for secure storing and sharing of the sensor data. We consider four entities in the system: IoT sensors, data requester, cloud provider, and the blockchain, as shown in Figure 1.

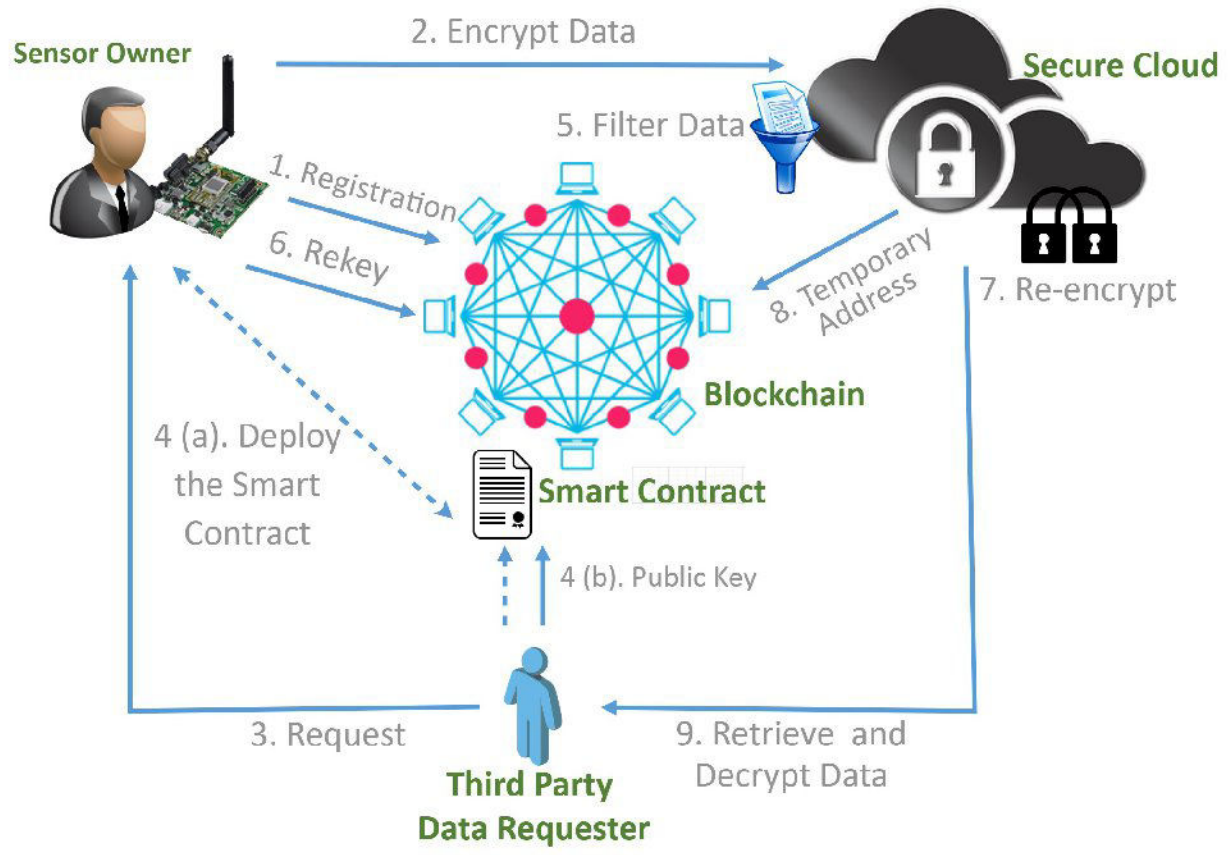

Fig. 1: Proposed Architecture

1) The sensors' owner activates the sensors, and registers them on the blockchain via a smart contract function.
2) After successful registration, the sensors' owner provides the sensor with the required key material such that the measured data can be sent encrypted to the cloud storage server.
3) A user requests access to one (or a group of) sensor(s) of the owner via the smart contract function.
4) After receiving the request, the sensors' owner and requester come to an agreement, a smart contract is generated and mined on the blockchain. The requester interacts with the blockchain to share the public cryptographic key and manages all the financial associated transactions.
5) On receiving the user request, cloud storage is notified by the blockchain. The software then filters the data according to the request.
6) Re-encryption cryptographic key from the sensor owner is updated on the smart contract when the user request is received.
7) Cloud server decrypts and re-encrypts filtered data, before storing it again on a temporary location onto the cloud server.
8) The encrypted data is temporally stored on the server and a transaction containing the address of the stored data is mined on the blockchain.
9) When the data is ready, the requester is notified of the temporary location by the blockchain. The requester can decrypt the data using its private cryptographic key.

## IV. Security Aspects

We propose to apply a Certificate Based Proxy Re-Encryption (CB-PRE) scheme, which constitutes of seven polynomial-time algorithms: Setup, CertifiedUserKeyGen, Encrypt, ReKeyGen, ReEncrypt, Decrypt1, and Decrypt2. We now explain each of these phases into more detail. In our proposal, we have combined the phases UserKeyGen and Certify to one phase called the CertifiedUserKeyGen phase.

- SetUp($l$): Given a certain security parameter $l$, the following steps will be executed to derive the public parameters $params$ and the master secret key $msk$.
  - First, the CA chooses an $l$-bit prime $q$. Next, an EC of order $q$ is generated, and a corresponding generator point $P$ is defined. Denote by $G$ the group of EC points.
  - A random value $\alpha \in F_q^*$ is chosen and $P_\alpha = \alpha P$ is computed.
  - Four different hash functions are determined. $H_1 : G \times \{0,1\}^{32} \rightarrow F_q^*$, $H_2 : F_q^* \times \{0,1\}^{64} \rightarrow F_q^*$, $H_3 : \{0,1\}^{64} \times G \rightarrow F_q^*$, $H_4 : F_q^* \times \{0,1\}^{64} \times \rightarrow F_q^*$.
  - The public parameters are now $params = \{G, q, P, P_\alpha, H_1, H_2, H_3, H_4\}$ and the master secret key is put as $msk = \alpha$.
- CertifiedUserKeyGen($params, id_U$): This algorithm is based on the Elliptic Curve Qu Vanstone (ECQV) certificate mechanism [33] and consists of the following three phases:
  - First, the involved entity $id_U$ generates a random value $r_U \in F_q^*$ and computes $R_U = r_U G$. Next the tuple $(id_U, R_U)$ is sent to the CA.
  - Upon arrival, the CA checks the identity of $id_U$. Next, it also chooses a random value $r_t \in F_q^*$ and computes $R_t = r_t P$. Then the certificate $Cert_U = R_U + R_t$ is derived. Finally, auxiliary information to derive the private key for the involved entity is computed by $r_a = H_1(Cert_U \| id_U) r_t + \alpha$. The tuple $(r_a, Cert_U)$ is sent back.

– The involved entity computes first its private key $d_U = H_1(Cert_U \| id_U) r_U + r_a$. Its public key equals to $P_U = d_U P$. If $P_U = H_1(Cert_U \| id_U) Cert_U + P_\alpha$, it accepts the key pair $(d_U, P_U)$.

- Encrypt($params, M, id_A, d_A, T_0$): The metadata is generated for the message $M$, ie. $meta = (id_A \| T_0)$. Next, the following computations are made.

$$
\begin{aligned}
r &= H_2(d_A \| meta), R = rP \\
C_A &= M \oplus H_3(meta \| rP_A) \\
h_A &= H_4(C_A \| meta) \\
s_A &= r - h_A d_A
\end{aligned}
$$

The output $C$ of this algorithm equals to $C = (C_A, meta, h_A, s_A)$.

- ReKey($params, d_A, id_B, Cert_B, C_A, meta$): First $r = H_2(d_A \| meta)$ is derived from $C$. Then, the public key of $id_B$ is computed as $P_B = H_1(Cert_B \| id_B) Cert_B + P_\alpha$. This leads to the definition of the ReKey as

$$
rk_{AB} = H_3(meta \| rP_A) \oplus H_3(meta \| rP_B)
$$

The output is the key $rk_{AB}$.

- ReEncrypt($params, C_A, rk_{AB}$): The re-encryption phase changes the ciphertext $C_A$ to $C_B$ by

$$
C_B = rk_{AB} \oplus C_A
$$

Note that $C_B$ also corresponds to $M \oplus H_3(meta \| rP_B)$, which will be used in the decrypting phase of the delegate. The output $C'$ is now the tuple, containing $C_B, meta, ID_B, h_A, s_A$.

- Decrypt1($params, C, d_A$): Here the delegator wants to decrypt the ciphertext to derive the original message and to check its authenticity. Therefore, the following computations are required:

$$
\begin{aligned}
r &= H_2(d_A \| meta) \\
M &= C_A \oplus H_3(meta \| rP_A) \\
h_A &= H_4(C_A \| meta) \\
\text{Check: } s_A &= r - h_A d_A
\end{aligned}
$$

- Decrypt2($params, C', d_B$): In this phase, the delegate $B$ derives the message $M$ from $C'$ by the following operations.

$$
\begin{aligned}
R &= s_A P + h_A P_A \\
M &= C_B \oplus H_3(meta \| d_B R) \\
\text{Check: } h_A &= H_4(C_A \| meta)
\end{aligned}
$$

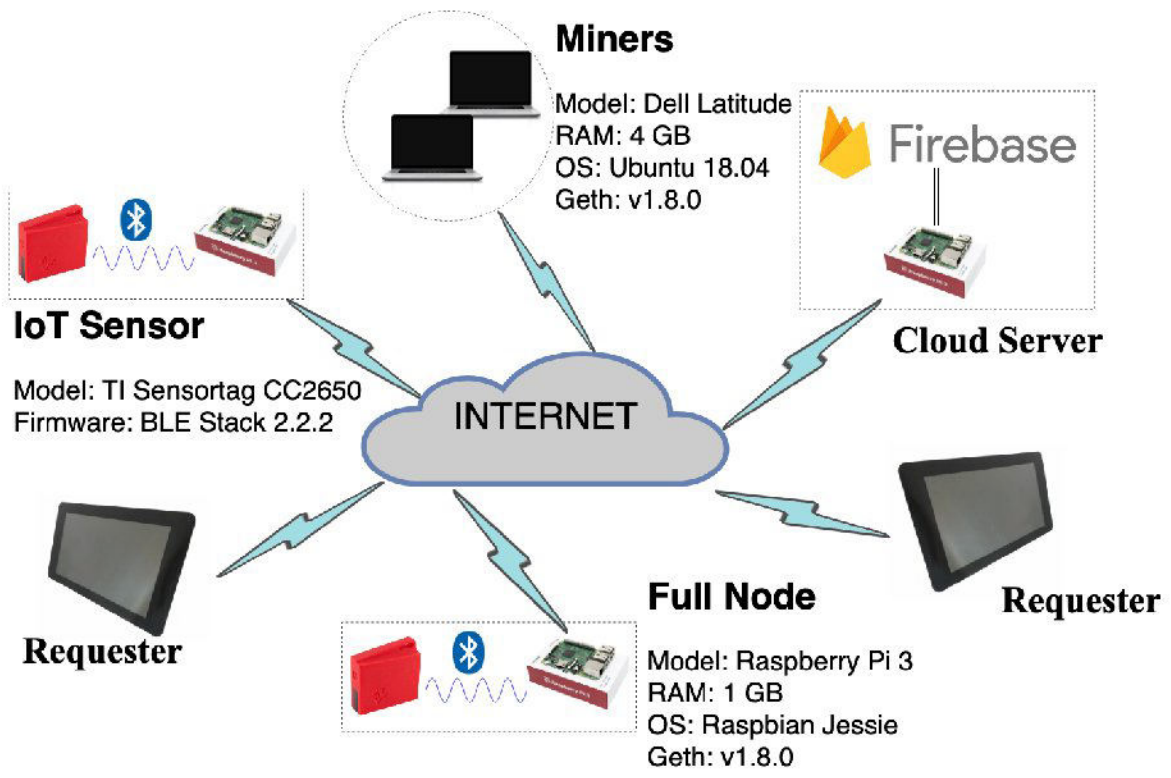

Fig. 2: Overview of the Architecture Implemented

## V. Implementation

We demonstrate the feasibility of the system design with the prototype implementation containing a permissioned Ethereum blockchain, IoT sensors and a cloud server for storage of the data. Figure 2 illustrates the setup of the system with three IoT sensors, three mining computers, five ethereum full nodes, two regular users and one cloud storage server.

We configured and connected all the devices to the internet. We used the auto-discovery protocol of Geth to connect the miners and the full nodes, and configured google firebase cloud for storage.

### A. Miners

The proposed system consists of three miners that generate a block of transactions on average every 13 seconds. These miners are running on a virtual machine with the same hardware capabilities. All the mining devices were configured to use one Ethereum wallet that collects the mining reward. These miners are running on Geth v1.80 [34] with four mining threads each.

### B. Smart Contracts

We developed two smart contracts[1] on truffle [35] and compiled them with Solidity 0.4.24 [36]. The first smart contract consists of the functions to register the sensor, request data, and financial functions. The second smart contract is dynamically created in the runtime when the user requests for the data.

### C. IoT Sensors

Each sensor TI Sensortag CC2650 connects to a Raspberry Pi 3 Model B (RSP) through Bluetooth Low Energy, as shown in Figure 2. This RSP manages the sensor and the Ethereum account to perform transactions on the blockchain on behalf of sensors. A sensor application is developed in Python 2.7.12 that connects to the sensor, performs the cryptography functions described in the proxy re-encryption scheme on the sensor data and uploads that data to the cloud storage server. This application synchronizes with the blockchain using the Python- JSON-RPC (JavaScript Object Notation - Remote procedure Calls) library. The MAC address of each sensor acts as its identity and is used for re-encryption. Once registered, the sensor starts uploading the encrypted data to the cloud server. It is assumed that the BLE connection between sensor and RSP is completely secure.

[1] https://github.com/ahsan100/smart-contract

### *D. User Application*

A customized application is designed as the user interface in Python 2.7.12, running on a Raspberry Pi 3 attached to a touchscreen. This application uses JSON-RPC to get the sensors' information from the blockchain. After selecting the required sensor, the user enters details for specifying the data requirements. We deploy a new smart contract on the blockchain in run-time based on the user-selected options for the requested data (e.g. Sensor selection, Price). This application keeps track of the Ethereum wallet along with ECC [37] secret key of the data requester. The application downloads the data from the cloud server, checks for the signature and integrity, and decrypts the requested data.

### *E. Cloud Storage Server*

The cloud storage server consists of the RSP and the Google Firebase. RSP acts as ethereum full node and connects to the blockchain, while Google Firebase is used for the storage of the data. The authentication and integrity of the data are performed on the RSP and encrypted sensor data along with the meta-data is upload to the Google Firebase in JSON format. This cloud also performs proxy re-encryption and updates the smart contract variable for data address sharing.

## VI. Performance analysis

In this section, we describe the experiments to evaluate the proof of concept implementation. Experiments were designed to study the performance of the framework. We have performed multiple experiments to test the impact of proxy re-encryption on the overall system and performed some scalability tests.

### *A. Impact of Proxy Re-Encryption*

In the first experiment, we measure the impact of proxy re-encryption on the proposed system. The sensor encrypts the data before uploading it to the cloud storage and later re-encrypts it for sharing the data.

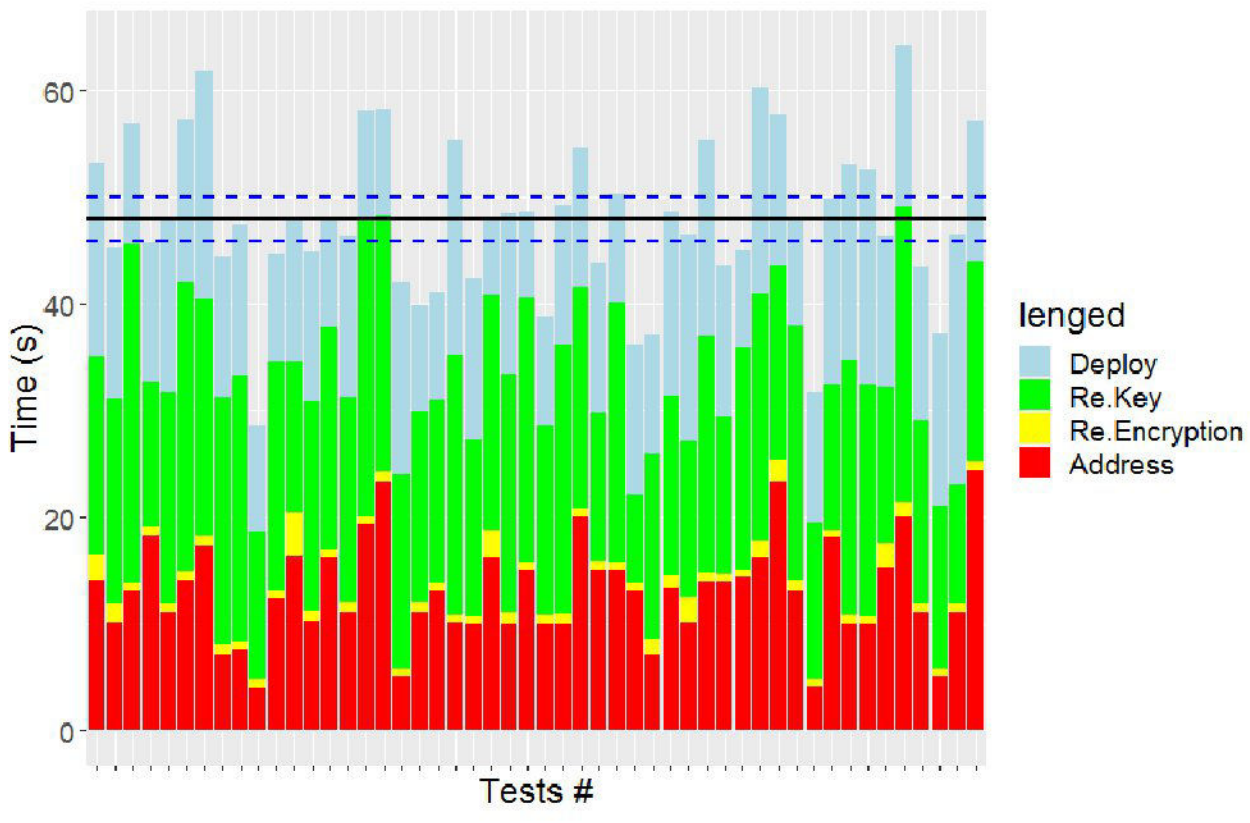

Fig. 3: Impact of Proxy Re-Encryption

In Figure 3, the time of the different parts in the scheme is illustrated. As can be seen, it takes on average 48.01 s to share the encrypted data with the user after the initial request with a confidence interval of 2.07 s. Consequently, adding proxy re-encryption to the scheme increases the delay by 60% due to the mining of the re-encryption key.

### *B. Scalability*

In the second experiment, we measure the scalability of the architecture by performing multiple transactions from multiple requesters to the sensor. The whole process was repeated 10 times for each scenario before taking the average. In the first scenario, only 1 request was initiated by the user and time was measured from the request to the retrieval of data by the requester. In the latter scenarios, the process was repeated by increasing five requests until the overall request reached 50.

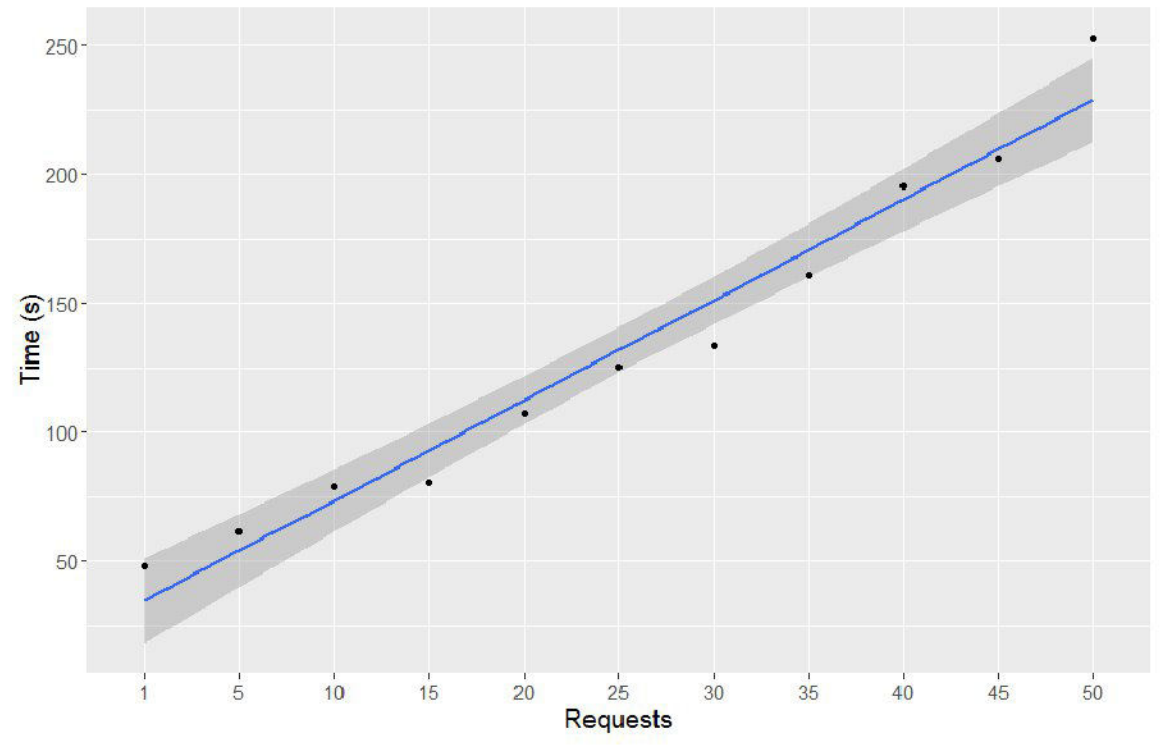

Fig. 4: Scalability Test

As seen from the Figure 4, the process shows a gradual increase in the delay due to the increase of transactions. This increase in the delay is caused by the scalability problem of the Ethereum blockchain. There seems to be a tradeoff between speed and reward for the generation of the new block. The number of transactions mined in a single block of Ethereum blockchain depends on multiple factors such as gas price and limit.

## VII. Conclusions

In this paper, we have proposed a blockchain based trading platform with the combination of a paring free proxy re-encryption scheme to ensure secure transfer of the sensor data to the user. We have also validated the proof of concept model on a private Ethereum testbed and demonstrated the practicality of the system design using off-the-shelf laptops and raspberry pis.

In the future, we plan to extend the proposed system with an implementation on a different blockchain platform e.g. Hyperledger. We also plan to extend our architecture by adding a distributed cloud storage to make the system more scalable.

## Acknowledgement

This work has been performed under the framework of SECUREConnect, 6Genesis Flagship (grant 318927), RESPONSE 5G (Grant No: 789658), CA15127 (RECODIS) and CA16226 (SHELD-ON) Projects.